\pdfoutput=1
\documentclass[
	a4paper, 
	10pt, 
	unnumberedsections, 
	twoside, 
]{LTJournalArticle}

\usepackage{tabularx}
\usepackage{cuted}
\usepackage{caption}
\usepackage{float}
\runninghead{} 
\footertext{\textit{Preprint version}} 
\title{LLM-Augmented Flow for Project Conformation and Task Complexity Evaluation Based \\on the TLM Framework} 

\author{%
	Maksim Sysoev\textsuperscript{1}\thanks{Corresponding author: \href{mailto:maksim.andr.sysoev@gmail.com}{maksim.andr.sysoev@gmail.com}}, Alexander Semenov\textsuperscript{2}, Ilia Cheshev\textsuperscript{2}, Alexey Gevorkyan\textsuperscript{2}, Danil Somov\textsuperscript{1}
}

\date{\footnotesize\textsuperscript{\textbf{1}}Independent Researcher, Thailand.\\
\textsuperscript{\textbf{2}}Independent Researcher, Cyprus.}

\renewcommand{\maketitlehookd}{%
	\begin{abstract}
		\noindent The software development process, from inception to a working product or solution, can be divided into two major phases: discovery (Discovery) and delivery (Delivery). The Discovery is further subdivided into High-Level Discovery (HLD) and Low-Level Discovery (LLD). During HLD, management identifies the business solution or product to be integrated into the company's infrastructure. LLD, in turn, produces requirement documentation artifacts, including the Product Requirements Document (PRD), Business Requirements Document (BRD), and Functional Specifications Document (FSD). Story Points (SP) are a widely used modern approach to measuring the effort required to accomplish predetermined tasks, and the artifacts listed above help derive SP more precisely. With the rapid growth of LLM technologies, one of the main questions that arises is whether the whole Discovery phase, or most of it, can be completed with the help of LLM tools.
        \\ In this paper, we test a framework that structures requirement documentation generation and task complexity scoring, based on the model introduced by \parencite{Tran2004} and executed with the help of agentic tools. The approach is applied to a real-world commercial project in the fintech domain as a feasibility study rather than a validation: development effort was not recorded for the tasks examined, and the resulting complexity scores are therefore reported as the output of the instrument rather than as validated estimates of effort.
	\end{abstract}
}

\begin{document}
\maketitle 

\section{Introduction}
In many organizations, business departments identify the features and products that need to be developed in order to achieve specific business objectives. Regardless of team structure or organizational hierarchy, delivering a new feature or product requires a clear idea of what is to be built. For this purpose, managers or product owners typically produce thorough, already low-level documentation tailored to their domain – PRD. As \parencite{Gohlich:2021} stress, each industry adopts its own approach to authoring PRDs which makes this type of artifact highly diverse.

The understanding of a BRD varies considerably across the literature, with no single universally accepted definition of its scope or purpose. Some guidelines describe the BRD as a document authored by senior management to calibrate and lead the company's strategic direction across short- and long-term horizons. Others maintain that the BRD should specify the precise business logic required to satisfy the product requirements outlined in the PRD. In this paper, the BRD is treated as a knowledge asset that captures the business algorithms and processes necessary to achieve the milestones set by management. Once the PRD is handed down to mid-level managers and business analysts, the natural next step is the derivation of a BRD grounded in the product requirements.

The rules and approaches for generating FSDs have been studied in the academic literature for several decades. \parencite{Parnas1995} treat FSD — referred to in their work as a System Requirements Document (SRD) — as a specification that regards the system as an unknown, opaque structure to which mathematical models and formally quantifiable logic should be applied. Similarly, \parencite{Kim2005} adopt a program functional specifications approach that also relies on measurable attributes, expressed in clear and intuitive natural language. In this paper, the FSD is treated as a structured set of requirements underpinned by formal logic, incorporating illustrative examples of code snippets, data models, API requests, metadata, and other related artifacts. Within this framework, the FSD is regarded as the highest-priority document in the discovery pipeline, despite the fact that it is derived from the BRD as its foundational input.

In this paper, an LLM is used to generate all three documents. The PRD is produced through free-form prompting, whereas the BRD and FSD are produced with dedicated LLM skills. The two differ in the degree of human involvement: the Business Rules of the BRD are authored manually by an analyst and serve as the input from which the LLM derives the Business Requirements, whereas the FSD is generated by the LLM in full and adjusted only after review. This distinction is described in detail in the Materials and Methodology section.

To support real-time visual collaboration during the committee session, SaaS whiteboard tools are commonly employed. The project manager, holding a comprehensive picture of the work required, leads the process of structuring the solution into a coherent set of tasks. Upon completion, the full scope is composed into a sequential pipeline of tasks with clearly defined interdependencies. As a final step, the committee assesses the effort required to complete each task. Story points are widely used across the industry as a relative measure of task effort. However, as \parencite{Porru2016} note, human story point estimates tend to be overly optimistic, susceptible to cognitive bias, and vulnerable to social influence — particularly when managerial presence affects the scoring dynamic. In this paper, a different approach is adopted: the committee may optionally assign estimated hours to each task, while an LLM subsequently derives a corresponding complexity score — the analogue of story points (SP) — once the tasks have been created.

As \parencite{DaSilvaNeo2025} note, story points do not directly measure the time required to complete a task; rather, they estimate task complexity, volume of work, and risk — factors that closely correlate with the notion of a complexity score adopted in this paper. Alongside this, \parencite{Porru2016} observe that Planning Poker, using Fibonacci numbers as scoring options, is a widely adopted human estimation technique. Fibonacci numbers are employed in this context, as \parencite{DaSilvaNeo2025} explain, to discourage overly precise estimation and instead encourage the use of relative measurement markers. In other works exploring LLM-based project effort estimation, such as \parencite{Islam2025}, authors rely on historical datasets of user stories with pre-assigned story points as the basis for model training and evaluation.

\begin{figure}[htbp]
    \centering
    \includegraphics[width=0.5\textwidth]{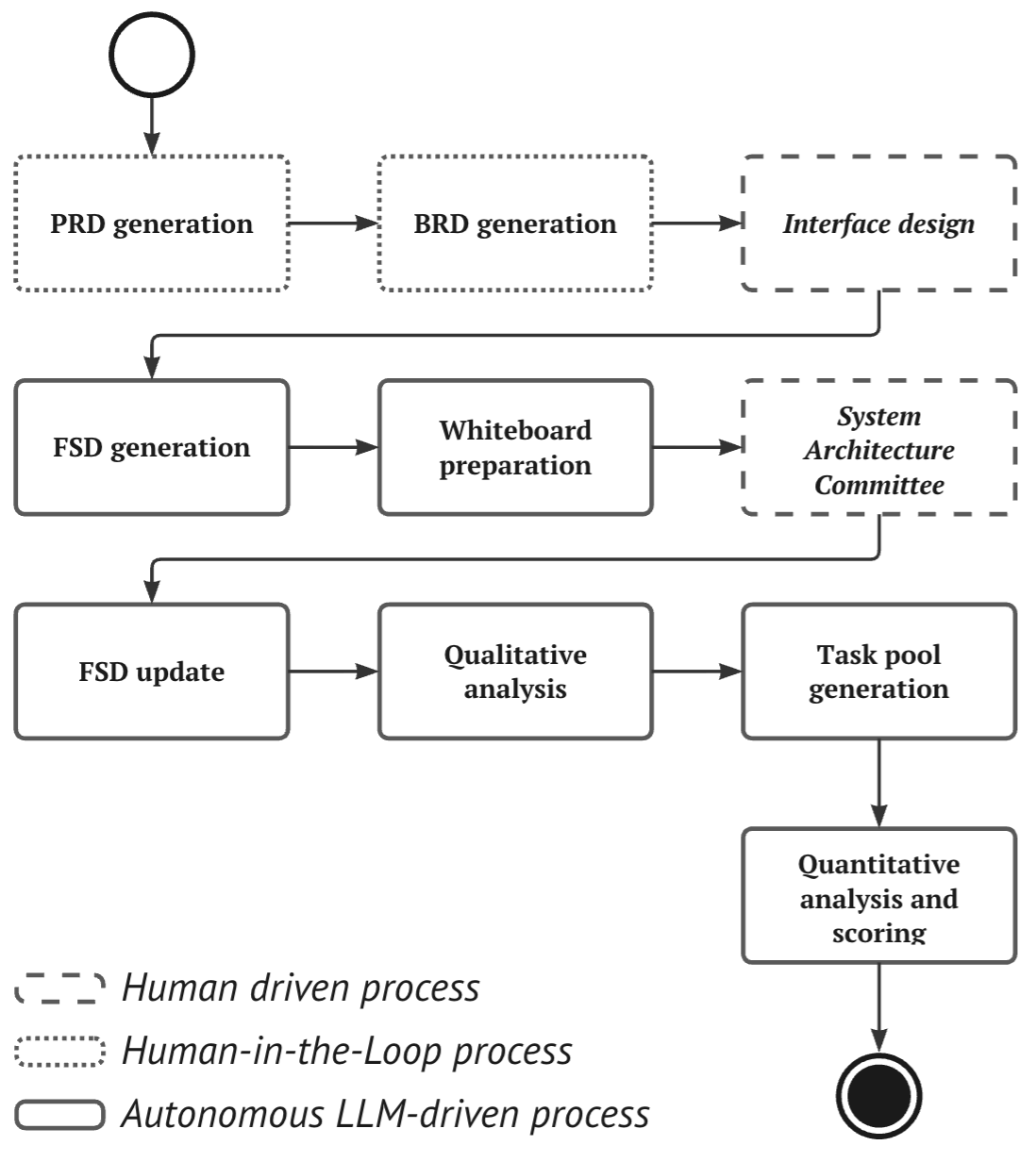}
    \caption{Pipeline of the LLM-augmented project compilation process, from requirement documentation generation to task creation and complexity scoring.}
    \label{fig:pipeline}
\end{figure}

The complexity score proposed in this paper differs fundamentally from both of these approaches: it neither employs Fibonacci scales nor depends on historical estimation data. The complexity score evaluation comprises two components: qualitative and quantitative. The qualitative component consists of metrics scored on a scale from 1.0 to 3.0, where the LLM is guided by reference examples to assess and reason about each criterion. For the quantitative component, the complexity measurement framework introduced by \parencite{Tran2004} — hereafter referred to as the TLM framework — is applied.

Theoretically, the majority of the proposed pipeline (Figure \ref{fig:pipeline}) is eligible for LLM-driven automation, with the exception of certain stages that inherently benefit from human judgement. In this paper, the "system architecture committee" stage is intentionally kept manual, as it serves as the primary checkpoint for human oversight, refinement, and final approval of the project scope. Certain other processes similarly retain a human-in-the-loop character by design. The quantitative and qualitative complexity scoring of tasks, however, is entirely delegated to the LLM.

In this study, all evaluations were conducted on real-world cases rather than synthetic data or simulations. Given the limited body of literature addressing this specific problem, no established methodology for generating representative mock data was identified in the open source literature. As a consequence, no machine learning mechanisms are employed in this work, since the absence of sufficient historical data precludes both model training and statistically robust validation. All experiments are therefore carried out exclusively on real undisclosed data.

The contribution of this paper is threefold. First, it specifies an LLM-augmented Discovery pipeline and its decomposition into discrete, individually invocable skills, each bound to a defined stage of requirement documentation generation and task composition. Second, it operationalises the TLM framework of \parencite{Tran2004} into a per-task score computable from documentation artefacts, extended with two competence-based qualitative dimensions grounded in the market- and resource-based views of the firm. Third, it measures the execution cost of running the pipeline in full on a real project, reported per skill in terms of time, token consumption and monetary expenditure.

The scope of the claims made here is correspondingly narrow. This paper does not claim that the complexity scores it produces are accurate predictors of development effort, nor that the pipeline reduces overall delivery time. However, the final motivation for the approach is indeed the reduction of the cost of generating Discovery artifacts and the mitigation of human bias in effort assessment. Thus, this paper registers the data acquired during the experiment for further research on the topic.

\section{Theoretical Framework for Task Complexity Evaluation}

\subsection{Theoretical Considerations}
The quantitative approach carries a clear axiomatic advantage over its qualitative counterpart: the application of mathematical models to numerical values produces results that are not subject to interpretive variation, unlike the qualitative reliance on broadly construed judgements such as ``good'' or ``bad'', whose meaning shifts according to the evaluator's background and values. \parencite{Almeida2017} corroborates this, noting additional advantages such as the relative ease of providing objective answers to conditional and hypothetical questions, the precision with which correlations between variables can be established, and the availability of a wide range of established techniques and statistical tests. However, the same authors acknowledge that the quantitative approach carries notable disadvantages: it requires a thorough prior understanding of the research domain, is prone to methodological weaknesses in survey-based designs, and faces challenges in replication due to the strict requirement for identical experimental conditions. \parencite{Heinrich2019} further identifies three compounding issues specific to quantitative research: intentional falsification of data, unintended measurement errors, and data incompleteness.

Two quantitative frameworks were considered for the measurement of task complexity. The first, proposed by \parencite{Tran2004}, derives three independent factors from the specification: the data groups involved, the conditions governing the logic, and the entropy of the resulting structure. The second, proposed by \parencite{McCabe1976}, proceeds differently, constructing a graph of the decision points within a program and their interdependencies and deriving a single figure from it.

The distinguishing consideration is the absence of an entropy term in the latter. Both frameworks employ graphs, but to different ends: \parencite{McCabe1976} uses the graph to calculate at the complexity figure itself, whereas \parencite{Tran2004} uses it solely to derive the entropy component, the remaining factors being counted directly. The entropy term is regarded as essential in the present work because it introduces a measure of structural disorder that a count of decisions alone does not capture. A single explicit figure carries the implication of a determinate difficulty, reproducible on each occasion the task is undertaken, whereas in practice no two executions of comparable work proceed identically, and the conditions under which a task is performed differ from one case to the next. The entropy component is intended to reflect that irregularity rather than to abstract it away. Thus, the TLM framework was adopted for this research.

For the qualitative approach, \parencite{Almeida2017} highlights a distinct set of limitations beyond subjectivity and evaluator bias, including ethical considerations, difficulties in generalising findings, and problems with the granularity of data. \parencite{Tariq2025} adds that the reliability of qualitative outputs is further contingent on the clarity of the research topic as perceived by the subject or object of study. \parencite{Zahle2025} stresses that insufficient description of the environment and infrastructure in which qualitative research was conducted can materially compromise the validity of its results.

The delegation of both analyses to an LLM is motivated by two considerations. The first is the automation of work that is otherwise manual and repetitive, namely the generation and adjustment of documentation artefacts and the derivation of scores from them. The second concerns the interpersonal dynamics of manual scoring: an evaluation performed by a model is not subject to deference towards managerial presence, nor to convergence on an estimate first expressed aloud by a senior colleague, and the criteria against which it scores are recorded in a file and are therefore inspectable and revisable, whereas the criteria applied by a human evaluator remain internal and largely unstated.

It would be mistaken, however, to characterise this as the elimination of bias. What the delegation achieves is the substitution of one bias profile for another. \parencite{Zhao2026} audit a range of biases in LLM-based evaluation across code generation, code repair and test generation, and find that several of them systematically shift the accuracy of the judgement, with the direction and magnitude of the shift varying by task and by model; verbosity in particular displays consistently negative effects, while the influence of others proves task-dependent rather than uniformly directional. Bias in this setting arises from at least three further sources beyond those identified in that work: the distribution of the training corpus, which conditions what the model treats as ordinary or exceptional practice and which is unevenly dense across technologies; the retrieval step, since scores that depend on retrieved context vary with what is retrieved for an otherwise identical specification; and the reference examples themselves, on which the model may anchor numerically rather than drawing only on the reasoning they are intended to convey.

The claim advanced here is accordingly narrower than the elimination of subjectivity. What the approach offers is consistency and auditability: the criteria are explicit, the approach is designed so that the same specification should yield the same score (a property not measured in the present study), and a scoring decision can be traced to the example that informed it. Neither property entails accuracy, and the reference examples remain the product of human judgement, such that any interpretive bias they carry is propagated systematically across every task scored against them rather than varying between evaluators.

Although \parencite{Yuan2023} demonstrate that models across a range of parameter counts are capable of performing elementary arithmetic tasks, all computation in the present approach is nonetheless delegated to Python scripts drawing on established mathematical libraries, with a view to the sustained application of the method rather than to any single execution. The intention is to eliminate the risk of miscalculation entirely: an arithmetic error committed by the model would not announce itself, since a plausible but incorrect entropy value is indistinguishable from a correct one without independent verification, and the cost of such an error compounds where scores are recorded against tasks and subsequently consulted. Deterministic computation removes this failure mode at negligible expense.

\subsection{Quantitative Analysis}
The theoretical foundation of the TLM-framework draws on two established bodies of work: the COSMIC \parencite{ISO19761} software measurement method and the task complexity model of \parencite{Wood1986}.

In software development, the primary deliverable is a software product comprising data and the functions that operate upon it. At the most fundamental level, all operations on data reduce to data movement and data manipulation. The COSMIC method classifies data movements into four types: entry — a data group received by the software from a user or external device; exit — a data group sent by the software to a user or external device; and read/write — a data group retrieved from or stored to a persistent storage medium. Each data movement handles exactly one data group at a time.

The TLM framework reconceptualises this structure as follows: software is defined as a set of functions, each comprising four fundamental elements — a stimulus (the trigger initiating the function), inputs, operation rules, and outputs.

While COSMIC decomposes the software product into its constituent elements, \parencite{Wood1986} addresses the complexity of the development task itself, identifying three principal complexity groups: component, coordinative, and dynamic. The TLM framework adopts the first two and explicitly excludes dynamic complexity, on the grounds that it is governed by temporal change. This exclusion is appropriate in the present context, where the majority of tasks are completed within one to four sprints — approximately 0.5 to 2 months. Dynamic complexity as defined by \parencite{Wood1986} would only become relevant during the business requirements evaluation phase, which falls outside the scope of this paper.

Component complexity ($TC_1$) is defined as a function of the number of distinct acts required to complete a task and the number of distinct information cues that must be processed in performing those acts:
\begin{equation}
    TC_1 = \sum_{p} \sum_{n} N_{ic}
    \label{eq:tc1}
\end{equation}
\noindent where $p$ is the number of distinct subtasks, $n$ is the number of distinct acts within each subtask, and $N_{ic}$ is the number of information cues associated with each act.

Coordinative complexity ($TC_2$) captures the complexity arising from precedence relationships between tasks — that is, the number of antecedent tasks that must be completed before a given task can be performed:

\begin{equation}
    TC_2 = \sum_{n} R
    \label{eq:tc2}
\end{equation}

\noindent where $n$ is the number of acts in the target task and $R$ is the number of precedence relations between the $i$\textsuperscript{th} act and all other acts.

\parencite{Tran2004} establish a correspondence between the COSMIC software model and the \parencite{Wood1986} task elements, as summarised in Table~\ref{tab:tlm_mapping}.

\begin{table}[ht]
    \centering
    \caption{Mapping between the COSMIC software model and \parencite{Wood1986} task elements.}
    \label{tab:tlm_mapping}
    \small
    \begin{tabular}{ll}
        \hline
        \textbf{COSMIC Model} & \textbf{Wood (1986) Interpretation} \\
        \hline
        Software System & Task \\
        Subsystem & Subtask \\
        Function & Act \\
        Input/Output group & Information cue \\
        \hline
    \end{tabular}
\end{table}

Building on this mapping, the TLM framework proposes three quantitative measures for software functional complexity, constituting a complexity tuple $(NOD,\ NOC,\ EOS)$.

\textbf{Number of Data Groups (NOD)} counts the distinct data groups — variables, structures, and analogous constructs — across all subsystems, corresponding to component complexity at the data level.

\textbf{Number of Conditions (NOC)} quantifies the internal branching complexity of each discrete function or subsystem. A standard two-way conditional (if/else) contributes 1 to the NOC; a three-way conditional (if/elif/else) contributes 2, and so forth.

\begin{figure}[htbp]
    \centering
    \includegraphics[height=4cm]{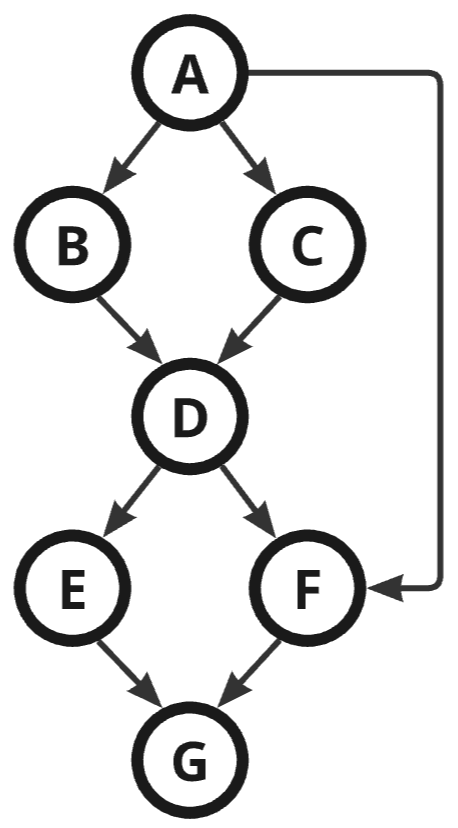}
    \caption{An example of a directed graph for EOS calculation.}
    \label{fig:graph_example}
\end{figure}

\textbf{Entropy of the System (EOS)} is derived by classifying all functional processes into equivalence classes based on their connectivity profiles: two processes are assigned to the same class if and only if they share identical in-degree and out-degree values, where degree is defined as the sum of arc weights entering or exiting the process node. This approach is motivated by the entropy-based software complexity measure proposed by \parencite{Davis1988}, who drew on the information entropy formula of \parencite{Shannon1948}. The EOS is computed as:

\begin{equation}
    EOS = -\sum_{n} P_i \cdot \log_2 {P_i}
    \label{eq:eos}
\end{equation}

\noindent where $n$ is the number of equivalence classes and $P_i$ is the probability of a functional process belonging to the $i$th class. In the implementation used in this study, EOS values below 1 are set to 1, so that tasks whose functional processes are structurally homogeneous retain their count-based magnitude rather than being reduced to zero. The quantitative complexity of a software task in the TLM-framework is thus expressed as the tuple:

\begin{equation}
    C = (NOD,\ NOC,\ EOS)
    \label{eq:cquant}
\end{equation}

To illustrate the EOS calculation, consider the directed graph in Figure~\ref{fig:graph_example} comprising seven nodes $A$ through $G$. Classifying each node by its in-degree and out-degree values yields five equivalence classes: $\{A\}$ with in-degree 0 and out-degree 3; $\{B, C, E\}$ with in-degree 1 and out-degree 1; $\{D\}$ with in-degree 2 and out-degree 2; $\{F\}$ with in-degree 2 and out-degree 1; and $\{G\}$ with in-degree 2 and out-degree 0. The resulting class size tuple is $(1, 3, 1, 1, 1)$ over a total of $n = 7$ nodes, giving probabilities $P_1 = \frac{1}{7}$, $P_2 = \frac{3}{7}$, $P_3 = \frac{1}{7}$, $P_4 = \frac{1}{7}$, $P_5 = \frac{1}{7}$. Substituting into Equation~\ref{eq:eos}:
\begin{equation}
    EOS = -\left(\frac{1}{7}\log_2\frac{1}{7} + \frac{3}{7}\log_2\frac{3}{7} + \ldots\right) \approx 2.128
    \label{eq:eos_example}
\end{equation}

\subsection{Qualitative Complexity Analysis}

The qualitative component of the complexity evaluation addresses contextual factors that cannot be captured by the quantitative TLM measures alone. While \parencite{Tran2004} provides a rigorous framework for structurally well-defined systems, it presupposes a thorough functional analysis with detailed specifications -- a condition that is not always met in agile-oriented organisations or those operating under resource constraints. In such contexts, developers must independently close contextual gaps, and the effort required to do so constitutes a meaningful dimension of task complexity.

To account for this, two competence-based measurements are proposed, grounded in the theoretical distinction between the Market-Based View (MBV) \parencite{Porter1980} and the Resource-Based View (RBV) \parencite{Wernerfelt1984, Barney1991} of the firm, as applied to knowledge management strategy by \parencite{Maier2001}. The distinction is employed here as a heuristic analogy at the level of the development team rather than as a formal transfer of the theory of the firm, whose justification at that level lies beyond the scope of this paper. Both measurements are assessed once per feature rather than per task: the factors are scored on the basis of the PRD, the BRD and the full task pool, and the resulting values are applied uniformly to the tasks derived from that feature. Each factor is scored on a continuous scale from 1.0 to 3.0, where the score represents the extent of contextual complexity and reference examples are provided to guide the LLM in calibrating its assessment through case-based comparison. A score of $1.0$ denotes the lowest contextual complexity, while a score of $3.0$ corresponds to the highest contextual complexity. 

All seven factors are defined to be monotonically increasing in complexity: for every factor, a higher score denotes a greater demand placed on the developer. This convention is a precondition for the aggregation defined in Equation~\ref{eq:m_competence} and Equation~\ref{eq:d_competence}, since factors of opposing polarity could not be averaged without transformation, and it removes a source of ambiguity in the reference examples on which the LLM's calibration depends.

\subsubsection{Market-Aligned Competence (MAC)}

MAC operationalizes the MBV dimension of task complexity, measuring the degree to which a developer must draw on external, market-derived knowledge to complete the task. Three factors are proposed: Stack Novelty (SN), which captures the degree to which the task requires adoption of technologies or architectural patterns that are new to the team's established stack; Solution Knowledge Scarcity (SKS), which measures the extent to which the market fails to provide reusable knowledge, documented patterns, and reference implementations for the problem class addressed by the task, such that the developer must derive an approach independently; and Known Vulnerability Exposure (KVE), which measures the degree to which the technologies or approaches required by the task carry known security risks documented in the broader market, such as publicly reported vulnerabilities.

\subsubsection{Domain-Aligned Competence (DAC)}

DAC operationalizes the RBV dimension, measuring the depth to which a developer must understand the organisation's unique internal resources --- its codebase, architectural decisions, undocumented workflows, and accumulated technical debt --- to complete the task successfully. Four factors are proposed: Legacy Burden (LB), which measures the degree to which legacy code constrains or complicates the task; Security and Data Integrity Sensitivity (SDIS), which measures the extent of domain-specific security knowledge and data handling diligence required; Business Process Depth (BPD), which measures the depth to which the developer must understand the organisation's specific business logic beyond what is explicitly specified; and Task and Documentation Context Deficiency (TDCD), which evaluates the extent to which the PRD, BRD, and FSD fail to furnish sufficient and unambiguous context for task execution, obliging the developer to reconstruct the missing context independently.

\subsection{Complexity Score Aggregation}

The quantitative and qualitative components defined above are combined into a single scalar score. Aggregation proceeds in two steps: the seven qualitative factors are first reduced to their two competence dimensions, and the resulting dimensions are then applied multiplicatively to the quantitative magnitude of the task.

Each competence dimension is taken as the arithmetic mean of its constituent factor scores:

\begin{equation}
    MAC = \frac{1}{3}\left(SN + SKS + KVE\right),
    \label{eq:m_competence}
\end{equation}
\begin{equation}
   DAC = \frac{1}{4}\left(LB + SDIS + BPD + TDCD\right)
    \label{eq:d_competence}
\end{equation}

The quantitative tuple of Equation~\ref{eq:cquant} is reduced to a magnitude term by summing its two count-based components and scaling the result by the structural entropy of the task. The competence dimensions then enter as multipliers, yielding the composite complexity score:

\begin{equation}
    C = \left(NOD + NOC\right) \cdot EOS \cdot MAC \cdot DAC
    \label{eq:complexity}
\end{equation}

The summation of $NOD$ and $NOC$ reflects the fact that both measure component complexity in the sense of \parencite{Wood1986} --- the former at the level of data, the latter at the level of control flow --- and are therefore treated as additive contributions to a single magnitude. $EOS$ scales this magnitude by the structural heterogeneity of the functional processes involved.

\begin{equation}
    C_{\mathrm{lite}} = \left(NOD + NOC\right) \cdot EOS
    \label{eq:complexity_lite}
\end{equation}

It must be emphasised that $C$ and $C_{\mathrm{lite}}$ do not lie on a common scale. The full form carries a multiplicative competence term, which the reduced form omits entirely; scores derived from the two forms are therefore not directly comparable, and any ranking that mixes them will systematically favour tasks scored under the full form.

The resulting score is not equatable with story points as conventionally practised; the analogy drawn in the Introduction concerns purpose rather than construction. Story points are assigned on a discrete and deliberately coarse scale --- most commonly the Fibonacci sequence --- precisely in order to discourage false precision, and their magnitude acquires meaning only relative to a team's own historical velocity. The score defined here is continuous, unbounded above, and derived from properties of the specification rather than from team history. 

\section{Materials and Methodology}
\subsection{Requirement Documentation Composition}

The approach proposed in this paper is applied within the context of a software development team operating in the fintech domain. All requirement documentation and task data used in this study therefore reflect domain-specific characteristics inherent to this field, including integrations with third-party providers and transaction processing.

\subsubsection{Product Requirements Document (PRD)}

The PRD is the least structurally constrained document among the three artifact types considered in this study. It imposes no prescribed section ordering, no mandatory formatting rules, and no strict content requirements. Nevertheless, certain informational elements are expected to be present: the business goals or problems the product is intended to address; the proposed solution and its key hypotheses; measurable metrics by which each hypothesis can be validated or rejected; an average user profile; and a high-level feature roadmap exposing the relative priority of planned features.

The PRD is generated by an LLM from a freestyle prompt provided by the responsible manager; no dedicated skills or tools were created for the purposes of this research. However, when a new product bears a close resemblance to an existing one — such as the onboarding of an additional payment provider — the PRD may be composed by referencing a prior document as a structural template, thereby accelerating the authoring process while preserving consistency across the knowledge base.

\subsubsection{Business Requirements Document (BRD)}

The BRD follows a more rigorous structure than the PRD, comprising several mandatory sections: Glossary, External References, Business Rules (BRs) and Business Requirements (BReqs). The second and third sections are filled in manually by an analyst. The BReq section is then filled in by an LLM agent with reference to the context in the BR section. The Glossary section is filled in last by the LLM agent, based on the terminology used in all other sections.

Each Business Rule (BR) is represented as a row in a two-column table structure, containing the following attributes: a summary of what is to be implemented; the expected behaviour expressed using the Gherkin Given/When/Then syntax; priority; the responsible stakeholder entity or department; and references to related BRs.

Each Business Requirement (BReq) represents a discrete unit of business logic and comprises a short description of the logic, a definition of done specifying the eligibility criteria for successful implementation, a set of use cases, and metadata including a unique identifier, priority level, and references to related BReqs. It should be noted that a single BR may be decomposed into one or more BReqs.

\subsubsection{Functional Specifications Document (FSD)}

The FSD is the most structurally rigorous document among the three, comprising the following mandatory sections: functional requirements, data requirements, interface requirements, non-functional requirements, requirement-task relations, and complexity evaluation.

Functional Requirements (FRs) specify the technical logic that must be realised within the codebase. The intended level of specification depth is sufficient for a mid-tier, non-flagship LLM to implement the requirement with a high degree of confidence. While no formal success threshold is targeted in this research and no established methodology is applied for this purpose, empirical practice suggests that thorough specification of elements such as API endpoint definitions — including request body, headers, URL structure, and response schema — alongside explicit conditional logic expressed as if/then constructs, returns the most unambiguous results. Each FR is annotated with metadata comprising: the responsible division (frontend or backend); a unique identifier; a definition of success; and a list of blocking dependencies on other requirements.

Data Requirements (DRs) specify the composition of a given data group, including field-level attributes such as data types, nullability, and default values, as well as the designated storage (e.g. cache, relational database, non-relational database). Where appropriate, DRs are supported with an SQL insertion example to unambiguously deliver the intended data structure to the developer or LLM. DR metadata includes: the responsible division; references to the Interface Requirements (IRs) and FRs that influence the DR; blocking intra-requirements dependencies; and performance expectations such as indexing rules and TTL configurations.

Interface Requirements (IRs) are derived from designs produced by the design team and are grounded in the BReqs, as the interface design itself is informed by the business requirements. Each IR specifies mandatory UI elements, motion and animation behaviours, navigation targets triggered by user interactions, and the colour palette applied to page elements.

Non-Functional Requirements (NFRs) define the performance characteristics intended to operate within nominal performance bounds, including metrics such as median bandwidth, uptime targets, and reliability expectations for external integrations.

Task Relations section consists of a mandatory linking section that establishes traceability between the FSD and the task management platform. The section is structured as a table in which each row corresponds to a task, referenced via an embedded hyperlink to the respective entry in the task management tool. The remaining columns enumerate the requirement identifiers — FRs, DRs, and IRs — that have been incorporated into the given task, thereby providing a clear and auditable mapping between documentation and implementation.

Complexity Evaluation is the section in which task complexity data is recorded. It comprises three subsections: Entropy of the System (EOS), Market-Aligned Competence (MAC), and Domain-Aligned Competence (DAC). The EOS subsection contains data about the total number of logic units, their grouping by compositional category, and the resulting entropy score. Both MAC and DAC subsections consist of a results block — containing numerical scores — accompanied by a qualitative explanation justifying each assigned score. The structural distinction between MAC and DAC lies in the number of evaluated factors each contains. The methodology underpinning the derivation of these scores is elaborated further in this paper.

\begin{table*}[t]
    \caption{Opus 4.6 vs Sonnet 4.6 benchmark comparison.}
	\label{tab:opus_vs_sonnet}
    \centering
    \begin{tabular}{lccc}
        \hline
        Benchmark & Claude Opus 4.6 & Claude Sonnet 4.6 & Difference (\%) \\
        \hline
        ARC-AGI v2 & 68.8 & 58.3 & 18.01\% \\
        BrowseComp & 84.0 & 74.7 & 12.45\% \\
        GDPval-AA & 160.6 & 163.3 & -1.65\% \\
        GPQA & 91.3 & 89.9 & 1.56\% \\
        Humanity's Last Exam & 53.1 & 49.0 & 8.37\% \\
        MCP Atlas & 62.7 & 61.3 & 2.28\% \\
        MMLU & 91.1 & 89.3 & 2.02\% \\
        MMMU-Pro & 77.3 & 75.6 & 2.25\% \\
        OSWorld & 72.7 & 72.5 & 0.28\% \\
        SWE-Bench Verified & 80.8 & 79.6 & 1.51\% \\
        Terminal-Bench 2.0 & 65.4 & 59.1 & 10.66\% \\
        \hline
        \multicolumn{3}{r}{\textit{Median difference}} & \textbf{2.25\%} \\
        \hline
		\multicolumn{4}{l}{\footnotesize Source: \url{https://llm-stats.com/models/compare/claude-opus-4-6-vs-claude-sonnet-4-6}} \\
    \end{tabular}
\end{table*}

\subsection{Tools and Infrastructure}
\subsubsection{Model Selection}

All experiments in this study were conducted using cloud-based models drawn from the standard Claude family (Opus, Sonnet and Haiku). Initial evaluations were carried out using the Claude Opus 4.6 model. However, it became evident during testing that the associated token costs were prohibitively high for the anticipated volume of discovery operations across future products. A comparative analysis of Claude Opus 4.6 and Claude Sonnet 4.6, drawing on benchmark data from benchlm.ai\footnote{https://benchlm.ai}, revealed that while Opus 4.6 achieves a benchmark score of 87 against Sonnet 4.6's score of 83, Sonnet 4.6 is approximately 40\% less expensive to operate. Furthermore, llm-stats.com\footnote{https://llm-stats.com/} reports a median performance difference of only 2.25\% between the two models based on the relative benchmarks (See table \ref{tab:opus_vs_sonnet}). The median is used as a single summary of the gap because the pipeline exercises heterogeneous capabilities, such as reasoning, retrieval, tool use and code generation, each of which is captured by a different benchmark.

The most notable performance gaps are observed in ARC-AGI v2, which evaluates abstract reasoning capabilities, and BrowseComp, which measures web search and information retrieval performance, with Opus 4.6 outperforming Sonnet 4.6 by 18.01\% and 12.45\%, respectively. This is of particular concern given that complexity metrics such as MAC and DAC rely heavily on reasoning, information retrieval and score elaboration. Accordingly, Sonnet 4.6 was selected as the working model for the pipeline. Its performance is lower than that of Opus 4.6 on several of the benchmarks considered, most notably those concerning abstract reasoning and information retrieval, but for the majority of the pipeline stages it is held to be sufficient for the tasks at hand at the present stage of development. Whether the reasoning-intensive stages --- in particular the qualitative competence evaluation --- would benefit materially from the more capable model remains an open question, and one that the present study does not attempt to settle.

\subsubsection{Context Sources}

In addition to the requirement documentation itself, the pipeline draws on an internal retrieval service over technical documentation, which supplies business-logic context, established practices and operational guardrails relevant to the feature under analysis. Documentation retrieval and task management operations are performed through the standard MCP integrations of the SaaS knowledge base (KB) and the SaaS project management and issue tracking (PMIT) platforms.

\subsubsection{LLM Skills Configuration}

Every LLM-driven and human-in-the-loop stage (except for PRD generation) has a dedicated skill assigned to it, as summarised in Table~\ref{tab:skills}. The skill prompts and reference examples are not published, as they are covered by confidentiality obligations; the structure and function of each skill are described below. Skills are stored following the conventional Claude directory structure: \texttt{.../skills/$\langle$skill\_name$\rangle$/SKILL.md}. Each skill contains a structured metadata header (frontmatter) specifying attributes such as name, description, compatibility, and related configuration fields, which are loaded at startup to inform the LLM of the available skills in its operational context. All skills that interact with the primary documentation artifacts — namely the BRD and FSD — also carry supplementary instructions for populating supporting sections of those documents, such as glossaries and task relation tables.

The \texttt{/brd-enrich} skill is responsible for finalising the BRD by deriving BReqs from the information present in the PRD and the pre-populated BRs in the BRD. Each generated BReq comprises a short description, a set of use cases, and a list of eligibility conditions. The \texttt{/fsd-compose} skill references the PRD for high-level product objectives, analyses the business logic and behaviour documented in the BRD, and consults the internal retrieval service for context on existing business and technical logic. Drawing on this context, it generates a structured set of FRs, DRs, and IRs (drawing also on design references, where available) in relation to the BReqs, including technical specifications and realisation examples — occasionally incorporating ready-to-use code snippets, endpoint definitions, and SQL insertion examples. The \texttt{/preload} skill contains instructions for grouping functionally related or jointly implementable requirements into cohesive tasks. Once a batch of tasks has been compiled, the skill creates a new board on the whiteboard (WB) SaaS platform via a custom-built toolset and populates it with visual task elements indicating the proposed task name and the responsible division.

The \texttt{/fsd-update} skill is optional in principle, though applied in the overwhelming majority of cases. Its purpose is to propagate the outcomes of the architecture committee review — which adjusts the order, count, and interdependencies of tasks on the WB — back into the FSD. Requirements rendered obsolete by these adjustments are not removed but are instead marked with strikethrough formatting to preserve the original information for reference. Requirements corresponding to newly introduced tasks are appended with a \texttt{new} tag and generated following the same procedure as in \texttt{/fsd-compose}, including context retrieval and BRD cross-referencing. The \texttt{/context-complexity} skill operates under the assumption that the PRD, BRD, and FSD are in an approved and finalised state. It parses all available documentation to evaluate and assign MAC and DAC scores, produces a qualitative elaboration justifying each score, and writes the results into the corresponding sections of the FSD. As reflected in Table~\ref{tab:skills}, this skill carries the largest prompt size among all skills in the pipeline, owing to the inclusion of reference examples that serve as the basis for case-based comparison. These examples guide the LLM in calibrating its scores and constructing the accompanying elaborations by providing annotated benchmarks against which each evaluated case is assessed. The \texttt{/compose} skill provides instructions for creating tasks on the PMIT platform and establishing the relational links between them. Finally, the \texttt{/valorization} skill is responsible for the quantitative analysis stage, evaluating NOD, NOC, and EOS values and writing the results into both the FSD and the relevant tasks on the PMIT platform.

\begin{table*}[t]
    \caption{LLM Skills Overview.}
    \label{tab:skills}
    \centering
    \begin{tabular}{llc}
        \hline
        \textbf{Skill} & \textbf{Executed at} & \textbf{Prompt Size} \\
        \hline
            \texttt{/brd-enrich}          & BRD generation                    & 3026 \\
            \texttt{/fsd-compose}         & FSD generation                    & 1939 \\
            \texttt{/preload}             & Whiteboard preparation            & 2291 \\
            \texttt{/fsd-update}          & FSD update                        & 3322 \\
            \texttt{/context-complexity}  & Qualitative analysis              & 7762 \\
            \texttt{/compose}             & Task pool generation              & 4241 \\
            \texttt{/valorization}        & Quantitative analysis and scoring & 4702 \\
        \hline
    \end{tabular}
\end{table*}

\subsection{Pipeline Workflow}

\subsubsection{Documentation Generation Stage}

The documentation generation stage is initiated during the HLD phase, in which business units deliberate on the intended solution and identify the most suitable product to achieve the defined business objectives. Following this, the product owner (PO) proceeds to author the PRD. The PO composes a custom prompt articulating the product expectations, target metrics, and other relevant considerations. This input is subsequently structured and automatically enriched by the LLM, which renders a preview for the PO's review. Upon approval, the LLM prepares the content in KB-compatible markdown format and publishes it to the corporate knowledge base via the default KB MCP.

The subsequent step involves the project manager and business analyst — consolidated into a single role in this study, hereafter referred to as BA — reviewing the PRD and initiating the creation of the BRD. The BA populates the BRD by authoring Business Rules (BRs), each of which establishes the semantic business logic context for subsequent LLM processing. The summary subsection of each BR captures the high-level business logic to be addressed, while the Given/When/Then section provides a thorough behavioural specification. The BA also appends relevant external documentation references. Once the BRs are in place, the BA invokes the \texttt{/brd-enrich} skill, supplying the KB product link and feature name as arguments. The skill first reads the PRD, then processes the BRs and proceeds to generate discrete BReqs mapped to every BR present in the document. Each BR must yield at least one BReq; a BR producing no BReq is treated as a skill execution failure. Upon completing BReq generation, the skill enriches the cloud-hosted BRD document with the generated BReqs.

Upon finalisation of the BRD, the BA invokes the \texttt{/fsd-compose} skill with the same KB product link and feature name as arguments. The skill begins by reading the BReqs to establish the business context and determine the logic to be realised. It then consults the internal retrieval service for domain-specific business logic, established practices and operational guardrails. Drawing on the full accumulated context, the LLM generates three categories of requirements — functional, interface, and data — and populates the corresponding empty FSD file on the feature page of the knowledge base.

The final step of the documentation generation stage involves preparing a whiteboard (WB) for the architectural committee review. The \texttt{/preload} skill, in conjunction with a set of dedicated Python scripts interfacing with the WB platform API, programmatically creates a new board named after the relevant feature and its parent product. The skill then populates the board with visually distinct colour-coded elements, each representing a proposed task. Task composition is determined autonomously by the LLM, which groups requirements that are functionally related or efficiently implementable together into a single task. Each task element is annotated with a proposed task name, the list of constituent requirement identifiers, and the responsible division.

\subsubsection{Postcommittee Stage}

Following the architectural committee session, in which tasks are reordered, dependency relationships are established, redundant tasks are removed, and any overlooked tasks are introduced, the \texttt{/fsd-update} skill is invoked. Due to the limited capabilities of the WB platform API, screenshots of the finalised task board are captured and placed in a designated project folder, serving as the visual input for the skill. The skill accepts three arguments: the KB link to the product, the feature name, and the project identifier — represented as an index in the current environment.

The skill begins with a holistic visual and contextual analysis of the board, identifying which tasks have been retained, modified, or newly introduced relative to the pre-committee state. Where new tasks have no applicable existing requirements, the skill generates the corresponding new requirements in the FSD file.

\begin{table*}[t]
    \caption{Execution metrics per discovery skill.}
    \label{tab:skill_metrics}
    \centering
    \begin{tabular}{lccccc}
        \hline
        Skill & Time* & Entries & Tokens & Cost (USD) & USD/100k tokens \\
        \hline
        /brd-enrich & 8:35 & 37 & 1,867,734 & \$3.03 & 0.1622 \\
        /fsd-compose & 14:32 & 46 & 3,343,337 & \$4.13 & 0.1235 \\
        /preload & 1:51 & 34 & 1,406,185 & \$1.13 & 0.0804 \\
        /fsd-update & 11:52 & 79 & 5,130,847 & \$4.49 & 0.0875 \\
        /context-complexity & 6:02 & 42 & 3,114,866 & \$2.65 & 0.0851 \\
        /compose & 11:40 & 56 & 3,963,018 & \$4.48 & 0.1130 \\
        /valorization & 7:13 & 49 & 3,208,439 & \$4.35 & 0.1356 \\
        \hline
        \textit{Total} & \textbf{61:45} & \textbf{343} & \textbf{22,034,426} & \textbf{\$24.26} & \textbf{0.1101}$^\dagger$ \\
        \hline
        \multicolumn{6}{l}{\footnotesize \textit{*Times given as mm:ss.}} \\
        \multicolumn{6}{l}{\footnotesize \textit{$\dagger$Overall value (total cost / total tokens).}} \\
    \end{tabular}
\end{table*}

Once the FSD has been adjusted and finalised, the \texttt{/context-complexity} skill is invoked. It performs a comprehensive scan of the FSD to establish a full understanding of the required implementation scope. DAC factors are assessed against the approved documentation and the retrieved internal context; MAC factors are assessed against the reference examples and general knowledge of the technologies involved. Upon completing the analysis, the skill writes the resulting scores and their accompanying elaborations into the dedicated sections of the FSD.

With the complexity evaluation in place, the \texttt{/compose} skill is invoked to handle task creation on the PMIT platform. It verifies that all relevant requirements are accounted for, creates the tasks via dedicated Python API tooling, and subsequently establishes the dependency relationships between them.

\subsubsection{Human Assessment Stage}
Once the pipeline had been executed in full, the same committee that had conducted the architecture review assessed the quality of the artefacts it had produced. The assessment was recorded on a five-point scale --- worse, bad, acceptable, good and excellent --- the levels of which are defined not by the perceived quality of the output alone, but by the remedial action each implies. \textbf{Worse} denotes an artefact offering no basis for improvement, requiring the approach itself to be reconsidered and the output reproduced from scratch. \textbf{Bad} denotes an artefact whose output is unsatisfactory while the underlying approach remains sound, such that the deficiency is expected to be remediable through adjustments to the framework rather than its replacement. \textbf{Acceptable} denotes an artefact adequate as a foundation, requiring improvement and refinement before use. \textbf{Good} denotes an artefact of high quality requiring only minor refinement, at a level nonetheless insufficient to dispense with human review. \textbf{Excellent} denotes an artefact of outstanding quality; it implies continued observation under the existing configuration and, should that standard be sustained across subsequent executions, consideration of the stage as a candidate for automated review with only occasional human oversight. Each reviewer assessed only those artefacts falling within their remit, which accounts for the absent entries in Table~\ref{tab:artefact_assessment}.

It is to be noted that the assessment was arrived at through internal committee discussion rather than independently by each reviewer. The procedure consequently admits the social influence effects that \parencite{Porru2016} identify in the context of manual estimation, and the unanimity of the assessment of the FSD may therefore reflect the course of the discussion in part, rather than four separately formed judgements. No measure of inter-rater agreement was computed, as agreement statistics of the conventional kind are not meaningfully estimable across four reviewers and a partially observed rating matrix; the ratings are reported as recorded.

\section{Data and Results}
\subsection{Case Study}

The original goal was to draw case study material from a range of forthcoming projects and products. The sample was, however, reduced to a single major project, which is presented here. This project was preceded by several smaller initiatives; these served as a testing and calibration ground rather than as eligible case material. The project selected is substantial in scope: the integration of a new third-party provider offering a broad range of options to users of the platform. A significant proportion of these options require additional input from the user, which in turn required additional frontend work — namely, the introduction of further modal windows and associated interaction logic.

\begin{table*}[t]
    \caption{Human assessment of LLM-generated Discovery artefacts by role.}
	\label{tab:artefact_assessment}
    \centering
    \begin{tabular}{lcccc}
        \hline
        Artefact & Reviewer A & Reviewer B & Reviewer C & Reviewer D \\
        \hline
        BRD & Good & --- & --- & --- \\
        FSD & Bad & Bad & Bad & Bad \\
        Whiteboard tasks & Acceptable & Bad & Bad & Acceptable \\
        Task composition & Good & Acceptable & Good & --- \\
        \hline
    \end{tabular}
\end{table*}

Although this was among the largest projects undertaken, it arrived at a point when the relevant parts of the system had already matured, which allowed the integration to proceed more smoothly than the preceding ones. The project was demanding in terms of the volume of information requiring analysis prior to development, rather than in terms of technical novelty. In the team's assessment, it was a generic product rather than a unique one: it did not call for novel approaches or architectural re-evaluation.

At a high level, development spanned backend integration with the provider, the corresponding frontend input logic and interface elements, and the processing of transaction outcomes.

It should be noted that actual development effort was not recorded for the tasks examined in this study. The team's task management configuration captures no logged hours, and no per-task hour estimates were assigned during the architecture committee session in this case; no retrospective reconstruction was attempted: post-hoc effort recall is subject to precisely the cognitive biases that the approach is intended to mitigate, and its use as a criterion would therefore be self-defeating. In consequence, the complexity scores reported in this paper are presented as the output of the instrument rather than as validated estimates, and no correlation against realised effort is claimed.

\subsection{Data}

As Table~\ref{tab:skill_metrics} shows, the composition of the FSD required the longest execution time, at slightly over fourteen minutes, whereas preparation of the whiteboard with the requisite task data was the fastest step, completing in just under two minutes. The FSD update step accounted for both the largest number of LLM entries recorded in a single session and the highest token consumption; the lowest token usage was again observed for whiteboard preparation. In terms of cost, the FSD update was likewise the most expensive skill, although it exceeded the second most expensive step by only one cent.

The assessments summarised in Table~\ref{tab:artefact_assessment} present a mixed picture. The BRD and the composition of task management entries were both rated positively, with the latter judged Good by two of the three reviewers who assessed it and Acceptable by the third. Whiteboard tasks divided opinion: two reviewers considered the output Acceptable, whereas the other two rated it Bad. The FSD, by contrast, was rated Bad unanimously by all four reviewers---the only artefact on which the assessment was both negative and uncontested. Read alongside the execution metrics in Table~\ref{tab:skill_metrics}, this result is notable: the two FSD steps together accounted for the greatest share of time, tokens, and cost in the pipeline, yet produced the artefact judged least usable, while the comparatively inexpensive whiteboard and task management steps received more favourable assessments.

\section{Discussion}

The pipeline was designed specifically to streamline the Discovery workflow, whose output serves as base material for the Delivery phase. Thus, a successful Discovery execution yields three outcomes: a set of supporting documentation, tasks prepared to a sufficient standard within the task management tool, and a workable development roadmap.

The rationale for augmenting the Discovery pipeline with LLMs is twofold: the artefacts must be readable by humans for review, and usable by LLM agents both as prompts and as an additional source of data. This would result in coding agents generating implementation code on the basis of the FSD and BRD, while also drawing on the roadmap set out explicitly on the whiteboard. Such agents could additionally report progress to users and reviewers by interacting with the task management system, closing tasks as the corresponding work is completed. For testing and QA agents, the same artefacts offer a valuable baseline: they establish what outcomes were originally expected, against which the agents' output can be evaluated during testing.

The pipeline therefore addresses three principal problems, all considered here within the scope of the Discovery phase alone. The first is speed: artefacts should be generated quickly enough that reviewers can reach them sooner and establish the state of the work with minimum delay. The second is quality, in the sense that human participants in the loop should not be required to make extensive corrections to the generated artefacts. The third is the cost of producing them.

On the first of these, a direct comparison against the manual process is not available, since no controlled baseline was recorded. In the team's experience, however, requirement documentation of comparable scope has typically required considerably longer to produce, against the sixty-two minutes of aggregate execution time reported in Table~\ref{tab:skill_metrics}. This contrast should be interpreted with care, and not as a straightforward measure of time saved.

Much of the manual process is consumed by clarification: the BA returns to the PO with questions the PRD does not answer, tech leads challenge assumptions embedded in the business rules, and the requirements are revised as these exchanges proceed. This iterative interrogation is a substantial contributor to the eventual quality of the documentation, and it is not reproduced by generating the same artefacts from a fixed input in a single pass. What the pipeline compresses is the drafting; what it does not perform is the questioning that manual drafting has by default.

The consequence is that clarification is displaced rather than eliminated, and is deferred to the review stage. The present results bear this out: the FSD, generated in approximately fourteen minutes and revised in a further twelve, was nonetheless assessed as unusable by all four reviewers, and the revision step alone recorded seventy-nine LLM entries --- substantially more than any other skill in the pipeline. The comparison is therefore not a manual multi-day process against one hour, but front-loaded clarification against one hour of generation followed by an unmeasured quantity of review and rework. 

As the results indicate, the highest cost per 100,000 tokens is incurred by the BRD generation step. In our interpretation, this reflects the fact that the step constructs the working context from scratch: it establishes the fundamental account of what is to be built, which subsequent skills then draw upon to produce further artefacts. FSD composition is approximately 24\% less expensive per token, although it remains costly in absolute terms. The whiteboard preloading step is the least expensive of the pipeline at roughly half the per-token cost of BRD generation, which we attribute to its reliance on context already established by preceding skills. On this reading, per-token cost is largely a function of how much context is available at the point a skill is invoked. The cost-tracking tooling, however, did not distinguish cached from uncached input tokens, so differences in cache utilisation between skills may account for part of this variation.

When absolute figures are considered, however, the FSD-related steps remain the most costly in both time and expenditure. This is attributable to the volume of rework required to bring the document to an acceptable level of reliability, and is consistent with the entry counts reported in Table~\ref{tab:skill_metrics}: the FSD update step recorded 79 LLM entries, substantially more than any other skill. The FSD was also assessed as the least reliable of the artefacts produced. In the reviewers' view, the underlying cause was insufficient guidance on which sources of context the model should treat as authoritative.

Two remedies were proposed. The first is to maintain a single comprehensive guidance file describing the structure of the context sources and identifying which of them are authoritative. The second is to annotate the sources themselves at a finer granularity, marking individual sections according to their current status. Either approach would require ongoing maintenance, but the second distributes that effort across the sources rather than concentrating it in a document that must be kept in step with them.

The valorization step was introduced to provide a clearer understanding of the effort required by individual tasks. Conventional story-point estimation, as commonly practised, is subject to human bias; the intention behind the valorization skill was to circumvent this and to arrive at an independent allocation of the effort a task genuinely demands.

A limitation of the data presented in Annex~A (Table~\ref{tab:addendum_valorization}) arises from the point at which it was collected. Metrics were gathered after the project had been implemented and released, by which time the task set had grown beyond its original composition: during development, engineers and business stakeholders introduced changes to the requirements for architectural and business reasons, and these changes generated additional tasks. The tasks carrying MAC and DAC values may therefore be read as approximating the scope foreseen by management and the development team at the planning stage, whereas those scored on the quantitative metrics alone largely correspond to work introduced subsequently. This distinction is incidental rather than by design, and the two groups should not be treated as a controlled comparison.

Notwithstanding this, the committee's informal, post hoc impression was that the ordering of the scores was broadly plausible: the highest-ranked task in Annex~A was also the task the team regarded as the most demanding of the project, and the lowest-ranked entries correspond to changes the team considered trivial. This impression is subject to the same retrospective bias noted in the Case Study section, and is reported as an observation rather than as evidence of validity. The resulting scores should nonetheless be treated as approximate. As noted above, the valorization and complexity skills received little calibration in the present case, and the reference examples on which they depend were rough initial estimates; the scores are therefore indicative of relative ordering rather than of absolute magnitude.

The data collected here will inform subsequent work aimed at improving the pipeline. The principal focus will be the FSD generation skill. As noted above, the immediate priority is better markup of the files the model draws upon, whether by means of a single comprehensive guardrail document or an alternative arrangement. A further avenue under consideration is the introduction of additional retrieval-backed agents specialised in specific areas of the system. Beyond this, effort will be directed towards automating the pipeline while retaining human-in-the-loop interaction at the point of final approval. The present study indicates that it remains premature to remove human oversight altogether if the artefacts are to reach the standard required to serve as a context base for the Delivery phase.

The principal limitation of this study is that it rests on a single execution of the pipeline against a single project. The metrics reported in Table~\ref{tab:skill_metrics} are therefore unreplicated point estimates rather than means, and no variance is available for any of them. Since inference in large language models is stochastic, and since the number of entries required to bring an artefact to an acceptable state depends on the reviewer as much as on the model, a second execution against the same inputs would not be expected to reproduce these figures exactly. Comparative statements drawn from them --- that one skill is more costly than another, or that per-token cost varies with available context --- should accordingly be read as observations from one run rather than as established relationships, and their confirmation requires repeated execution across multiple projects.

\printbibliography

\clearpage        
                  
\onecolumn

\section{Addendum}
\subsection{Annex A.}\label{sec:annexA}

\begin{table}[H]
    \centering
    \caption{Per-task complexity metrics for the case-study project, ranked by
             complexity score.}
    \label{tab:addendum_valorization}
    \footnotesize
    \begin{tabularx}{\linewidth}{@{}l >{\raggedright\arraybackslash}X c rrrrrr@{}}
        \toprule
        ID & Task & Div. & MAC & DAC & NOD & NOC & EOS & $C$ \\
        \midrule
        T01 & Large-scale integration of provider options
            & BE & ---  & ---  & 16 & 48 & 2.24 & 143.36 \\
        T02 & Core integration component
            & BE & 1.60 & 1.95 & 3  & 6  & 2.13 & 59.81 \\
        T03 & Form handling within the user flow
            & FE & 1.60 & 1.95 & 2  & 6  & 2.13 & 53.16 \\
        T04 & Endpoint adaptation for the new provider
            & BE & 1.60 & 1.95 & 3  & 5  & 2.13 & 53.16 \\
        T05 & Integration component for user input requirements
            & BE & 1.60 & 1.95 & 2  & 5  & 2.13 & 46.52 \\
        T06 & Dynamic-form component
            & FE & ---  & ---  & 16 & 6  & 1.92 & 42.24 \\
        T07 & Interface refinement
            & FE & ---  & ---  & 18 & 7  & 1.46 & 36.50 \\
        T08 & Extension of existing flows to the new provider
            & BE & ---  & ---  & 9  & 6  & 2.32 & 34.80 \\
        T09 & Provider client module
            & BE & 1.60 & 1.95 & 3  & 2  & 2.13 & 33.23 \\
        T10 & Configuration entity extension
            & BE & 1.60 & 1.95 & 2  & 2  & 2.13 & 26.58 \\
        T11 & Scheduled job adaptation
            & BE & 1.60 & 1.95 & 2  & 2  & 2.13 & 26.58 \\
        T12 & Scheduled job adaptation
            & BE & 1.60 & 1.95 & 2  & 2  & 2.13 & 26.58 \\
        T13 & Dynamic-form component for additional input
            & FE & ---  & ---  & 8  & 5  & 1.50 & 19.50 \\
        T14 & Adaptive calculation logic
            & FE & ---  & ---  & 11 & 7  & 1.00 & 18.00 \\
        T15 & State handling under new logic
            & FE & ---  & ---  & 8  & 3  & 1.58 & 17.38 \\
        T16 & Minor adjustment
            & BE & ---  & ---  & 6  & 2  & 1.00 & 8.00 \\
        T17 & Minor adjustment
            & FE & ---  & ---  & 5  & 3  & 1.00 & 8.00 \\
        T18 & Minor adjustment
            & FE & ---  & ---  & 5  & 3  & 1.00 & 8.00 \\
        T19 & Data entity extension
            & BE & 1.60 & 1.95 & 1  & 0  & 2.13 & 6.65 \\
        T20 & Form field enhancement
            & BE & ---  & ---  & 4  & 2  & 1.00 & 6.00 \\
        T21 & Minor adjustment
            & FE & ---  & ---  & 4  & 2  & 1.00 & 6.00 \\
        T22 & Minor adjustment
            & FE & ---  & ---  & 4  & 2  & 1.00 & 6.00 \\
        T23 & Minor adjustment
            & BE & ---  & ---  & 4  & 1  & 1.00 & 5.00 \\
        T24 & Form field enhancement
            & BE & ---  & ---  & 4  & 1  & 1.00 & 5.00 \\
        T25 & Administrative interface update
            & FE & ---  & ---  & 3  & 1  & 1.00 & 4.00 \\
        T26 & Minor adjustment
            & FE & ---  & ---  & 3  & 1  & 1.00 & 4.00 \\
        \bottomrule
    \end{tabularx}
 
    \vspace{0.5ex}
    \parbox{\linewidth}{\footnotesize
        BE = backend division; FE = frontend division.
        Rows carrying MAC and DAC values ($n = 9$) are scored as
        $C = (NOD + NOC) \cdot EOS \cdot MAC \cdot DAC$; rows marked ---
        ($n = 17$) received no competence assessment and are scored as
        $C_{\mathrm{lite}} = (NOD + NOC) \cdot EOS$. The two are therefore not
        on a common scale: each fully scored row carries a uniform uplift of
        $1.60 \times 1.95 \approx 3.12$ that no partially scored row receives,
        and the ranking should be read with this in mind.
        }
\end{table}

\end{document}